\documentclass[prl,nobalancelastpage,twocolumn,nofootinbib,superscriptaddress,showpacs]{revtex4-1}
\usepackage{color,amsthm,amsmath,amsfonts,graphicx,bm}
\usepackage{bm}
\usepackage{color}
\usepackage{amsfonts}
\usepackage{amssymb}
\usepackage{amsmath}
\usepackage{epsfig}
\usepackage{graphicx}
\usepackage{enumerate}
\usepackage{multirow}
\usepackage{verbatim}
\usepackage{color}
\usepackage{subfigure}

\begin{document}

\title{Loop optimization for tensor network renormalization}

\author{Shuo Yang}
\affiliation{Perimeter Institute for Theoretical Physics, Waterloo, Ontario, N2L 2Y5, Canada}

\author{Zheng-Cheng Gu}
\affiliation{Department of Physics, The Chinese University of Hong Kong, Shatin, New Territories, Hong Kong}
\affiliation{Perimeter Institute for Theoretical Physics, Waterloo, Ontario, N2L 2Y5, Canada}

\author{Xiao-Gang Wen}
\affiliation{Department of Physics, Massachusetts Institute of Technology, Cambridge, Massachusetts 02139, USA}
\affiliation{Perimeter Institute for Theoretical Physics, Waterloo, Ontario, N2L 2Y5, Canada}

\pacs{}

\begin{abstract}
We introduce a tensor renormalization group scheme for coarse-graining a two-dimensional tensor network that can be successfully applied to both classical and quantum systems on and off criticality. The key innovation in our scheme is to deform a 2D tensor network into small loops and then optimize the tensors on each loop. In this way, we remove short-range entanglement at each iteration step and significantly improve the accuracy and stability of the renormalization flow. We demonstrate our algorithm in the classical Ising model and a frustrated 2D quantum model.

\end{abstract}

\maketitle

\textbf{Introduction} In recent years, the tensor network (TN) approach \cite{Cirac:2009aa,Verstraete:2009aa} has become a powerful theoretical \cite{Cirac:2011aa,Perez-Garcia:2007aa,Schuch:2012aa,Schuch:2013aa,Pollmann:2009aa,Gu:2009aa,Chen:2010ab,Schuch:2010aa,Affleck:1987aa,Gu:2014aa,kraus:2009aa,Wahl:2013aa,Yang:2014aa,Yang:2015aa,Poilblanc:2015aa,Sahinoglu:2014aa,Bultinck:2017aa,Huang:2015aa,Huang:2016aa,Hauru:2016aa,Evenbly:2016aa,Czech:2016aa} and computational \cite{White:1992aa,White:1993aa,Nishino:2001aa,Maeshima:aa,Verstraete:2004aa,Vidal:2003aa,Vidal:2007aa,Orus:2008aa,Vidal:2008aa,Pfeifer:2008aa,Pirvu:2011aa,Pirvu:2012aa,Verstraete:aa,Levin:2007aa,Jiang:2008aa,Jordan:2008aa,Orus:2008aa,Orus:2009aa,Corboz:2010aa,Gu:2008aa,Gu:2009aa,Gu:2013aa,Xie:2009aa,Zhao:2010aa,Xie:2012aa,Banuls:2009aa,Lubasch:2014aa,Lubasch:2014ab,Phien:2015ab,Phien:2015aa,Pizorn:2011aa,Wang:2011ab,Evenbly:2015ac,Evenbly:2017aa,Bal:2016aa,Wang:2016aa,Genzor:2016aa,Evenbly:2016ab} tool for studying condensed matter systems. Many physical quantities, including the partition function of a classical system, the Euclidean path integral of a quantum system, and the expectation value of physical observables, can be expressed in terms of tensor networks. Evaluating these quantities is reduced to the contraction of a multidimensional tensor network. In the two dimensional case, many algorithms \cite{Verstraete:aa,Levin:2007aa,Jiang:2008aa,Jordan:2008aa,Orus:2008aa,Orus:2009aa,Gu:2008aa,Gu:2009aa,Xie:2009aa,Zhao:2010aa,Xie:2012aa,Banuls:2009aa,Lubasch:2014aa,Lubasch:2014ab,Pizorn:2011aa,Wang:2011ab,Evenbly:2015ac,Evenbly:2017aa,Bal:2016aa} have been developed to implement the approximate tensor contractions. Among these, the tensor renormalization group approach introduced by Levin and Nave \cite{Levin:2007aa} and its generalizations \cite{Jiang:2008aa,Gu:2008aa,Gu:2009aa,Gu:2013aa,Xie:2009aa,Zhao:2010aa,Xie:2012aa,Zhao:2016aa,Evenbly:2015ac,Evenbly:2017aa,Hauru:2016aa} have unique features: the tensor contraction is based on a fully isotropic coarse-graining procedure. Moreover, when applying the method to a system on a finite torus, the computational cost is lower than those based on matrix product states (MPS) \cite{Verstraete:aa,Orus:2008aa,Orus:2009aa,Banuls:2009aa,Lubasch:2014aa,Lubasch:2014ab,Pizorn:2011aa,Wang:2011ab}. 

However, the Levin-Nave tensor network renormalization (TRG, also referred as LN-TNR here) \cite{Levin:2007aa} is based on the singular value decomposition (SVD) of local tensors, which only minimizes the truncation errors of tree tensor networks. Several improvements \cite{Xie:2009aa,Zhao:2010aa,Xie:2012aa} have taken into account the effect of the environments, but they are still essentially based on tree tensor networks. These approaches cannot completely remove short-range entanglements during the coarse graining process. For example, in the 2D TN calculation of a partition function (or a path integral) TNR based on simple SVD cannot simplify the corner-double-line (CDL) tensor \cite{Levin:2007aa}, despite the CDL tensor describing a product state that should be simplified to a $1$-dimensional tensor. In Ref.~\cite{Gu:2009aa}, this issue was seriously discussed. The authors pointed out that to further remove short-range entanglement, it is crucial to optimize the tensor configurations that contain a loop. However, due to the computational cost, only a crude iterative method is used to implement the loop optimization strategy. We refer to that method as Gu-Wen tensor network renormalization (TEFR, also referred as GW-TNR here). Ref.~\cite{Gu:2009aa} showed that GW-TNR can simplify CDL tensors, resulting in a simple fixed-point tensor for gapped/short-range correlated phases. This led to the discovery of symmetry-protected topological (SPT) order. Recently, Ref.~\cite{Evenbly:2015ac,Evenbly:2017aa} introduced a method based on multi-scale entanglement renormalization ansatz (MERA) \cite{Vidal:2008aa} to completely remove short-range entanglement, even in critical systems. This approach is referred to as Evenbly-Vidal TNR (EV-TNR).

In this paper, we develop a new practical and accurate algorithm called Loop-TNR, which can optimize loop-like tensor configurations more effectively than GW-TNR. Loop-TNR can completely remove the short-range entanglement within a loop at each coarse-graining step, for both on- and off-critical systems. The performance of Loop-TNR is greater than EV-TNR, and it has a lower computational cost. To demonstrate this, we computed the central charge and scaling dimensions of the critical Ising model, and then examined the accuracy and stability of these data when undergoing coarse-grained transformations. All TNR methods can produce accurate central charge and scaling dimensions. However, their stabilities are significantly different. Loop-TNR and EV-TNR provide good stability (their data remain accurate after tens of iterations), while LN-TNR has the worst stability (its data remain accurate only for a few iterations). 

Our results suggest that all TNR approaches can produce a fixed-point tensor which appears as the low-index part of the tensor (with a proper choice of basis). The high-index part is not represented by the fixed-point tensor, and can be considered to be the ``junk'' part of the tensor. As we perform more TNR iterations, the junk part may grow and eventually destroy the fixed-point tensor at low indices. The accuracy of an algorithm represents the accuracy of the fixed-point tensor at low indices. Its stability represents the growth rate of the junk part of the tensor. We have found that Loop-TNR can significantly reduce the growth rate of the junk part. Moreover, Loop-TNR can be used to compute physical measurements of 2D   projected entangled-pair states (PEPS) with high accuracy.    


\textbf{Loop-TNR algorithm} The Loop-TNR algorithm has the same purpose as GW-TNR \cite{Gu:2009aa}; to eliminate local entanglement on a loop and determine the correct structures of fixed-point tensors. However, Loop-TNR significantly improves the numerical stability and accuracy of the renormalization group (RG) flow, especially for critical systems. The following illustrates the three main steps of the Loop-TNR algorithm. The first and last steps are exact, and the second is approximate. The method is discussed with regards to a square lattice, but  generalizations to other lattices are straightforward.

\begin{figure}[tbp]
\centering
\includegraphics[width=1.0\columnwidth]{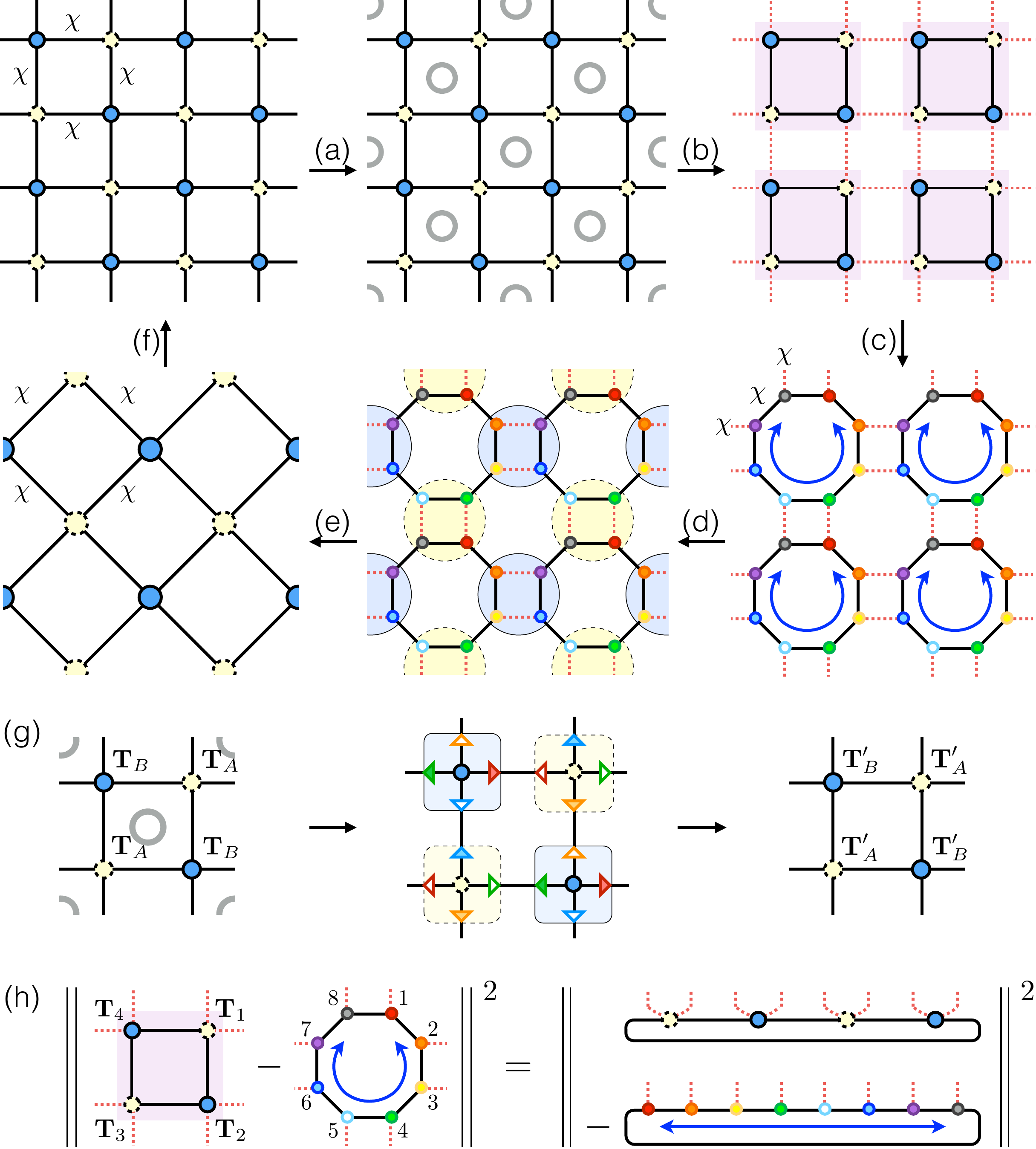}
\caption{(Color online)  
Three key steps of the Loop-TNR algorithm. (a) The entanglement filtering step. Projectors are inserted to eliminate local entanglements on the squares labeled with grey circles (see (g) for details). (c) The loop optimization step. Each of the shaded squares is deformed to a octagon made up of $8$ rank-$3$ tensors with bond dimensions no greater than $\chi$. The best approximation is found by minimizing the cost function in (h). (e) The same coarse graining step as in the standard LN-TNR algorithm. (h) The cost function of the loop optimization can be regarded as the distance between two MPS wave functions. The well-developed variational MPS method is applied to minimize the cost function.}
\label{fig:SquareNoSym}
\end{figure}

The Loop-TNR methods begins with an entanglement filtering step [Fig.~\ref{fig:SquareNoSym}(a) and (g)] with two important features. First, it provides a canonical gauge for every tensor, and filters out the local entanglement of off-critical systems. More specifically, two projectors are inserted on each bond shown in Fig.~\ref{fig:SquareNoSym}(g). These projectors are constructed in an iterative way based on QR decompositions \cite{Supplemental}. Subsequently, the tensors are redefined by combining the original tensors with the nearest projectors [see Fig.~\ref{fig:SquareNoSym}(g)] to complete the filtering step. In the Supplemental Materials, we show that this approach can completely remove the CDL tensors. Thus, for off-critical systems containing CDL tensors (with gauge transformations), our method can simplify the tensors and reduce the bond dimensions. Although there is no bond reduction in critical systems, the canonical gauge provided by this method can enhance the performance of the following step. This step is quite efficient because the overall computational cost scales as $\mathcal{O}(\chi^5)$, where $\chi$ is the bond dimension of the tensor.

In the next step the tensor network must be deformed from a square lattice to a square-octagon lattice [see Fig.~\ref{fig:SquareNoSym}(c)], as in the LN-TNR algorithm. However, approximations are necessary to avoid increasing the bond dimensions of the octagons. In the LN-TNR algorithm, this is achieved by minimizing the following single-site cost functions:
\begin{center}
\includegraphics[width=0.9\columnwidth]{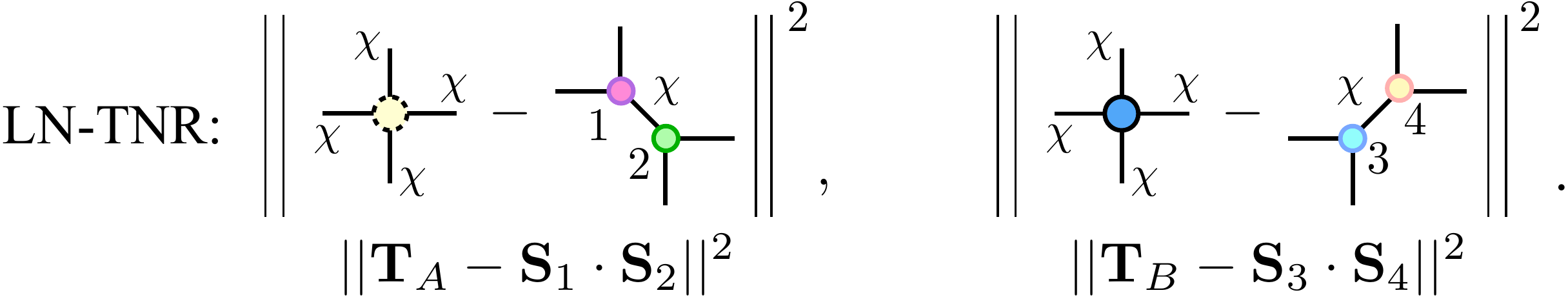}
\end{center}
The optimal $\mathbf{S}$ values are found using SVD and keeping only the largest
singular $\chi$ values. Here, ``$\cdot$'' means tracing over the indices of
connected bonds.

The Loop-TNR algorithm uses an alternative method to reduce the bond
dimensions. First, we define a cost function on the small patch shown in
Fig.~\ref{fig:SquareNoSym}(h), i.e.,
\begin{equation}
f=||\mathbf{T}_1 \cdot \mathbf{T}_2 \cdot \mathbf{T}_3 \cdot \mathbf{T}_4- \mathbf{S}_1 \cdot \mathbf{S}_2 \cdot \mathbf{S}_3 \cdot \mathbf{S}_4 \cdot \mathbf{S}_5 \cdot \mathbf{S}_6 \cdot \mathbf{S}_7 \cdot \mathbf{S}_8 ||^{2}
\label{cost}
\end{equation}
where the shaded square is deformed to an octagon. Since the cost function is now defined on a loop, we can remove the short-range entanglement inside this loop and significantly improve the accuracy, especially for critical systems. Furthermore, there is an efficient way to find the optimal $\mathbf{S}$ tensors by viewing each patch as a wave function made up of matrix product states (MPS) with periodic boundary conditions. The eight dotted lines shown in Fig.~\ref{fig:SquareNoSym}(h) are the physical legs of the MPS, and the solid lines are the virtual legs of the MPS. Minimizing the cost function is equivalent to minimizing the distance between two MPS. Thus, $\mathbf{S}$ tensors can be optimized using the well-developed variational MPS method \cite{Verstraete:2004aa,Verstraete:2009aa,Supplemental}. The computational cost of this step scales as $\mathcal{O}(\chi^6)$. The final step is the same as that of the LN-TNR algorithm. As shown in Fig.~\ref{fig:SquareNoSym}(e), a coarse-grained square lattice is obtained by contracting the tensor over the inner indices within the circles. The overall computational cost of all the steps only scales as $\mathcal{O}(\chi^6)$, which is significantly more efficient than other improved LN-TNR methods, such as SRG/HOSRG algorithms ($\mathcal{O}(\chi^7) \sim \mathcal{O}(\chi^{10})$ \cite{Xie:2009aa,Zhao:2010aa,Xie:2012aa}), and EV-TNR algorithms ($\mathcal{O}(\chi^7)$ \cite{Evenbly:2015ac} and $\mathcal{O}(\chi^6)$ \cite{Evenbly:2017aa,Note1}). Below, we demonstrate the advantages of the Loop-TNR algorithm using the classical Ising model on a square lattice.

\begin{figure}[tbp]
\centering
\includegraphics[width=1.0\columnwidth]{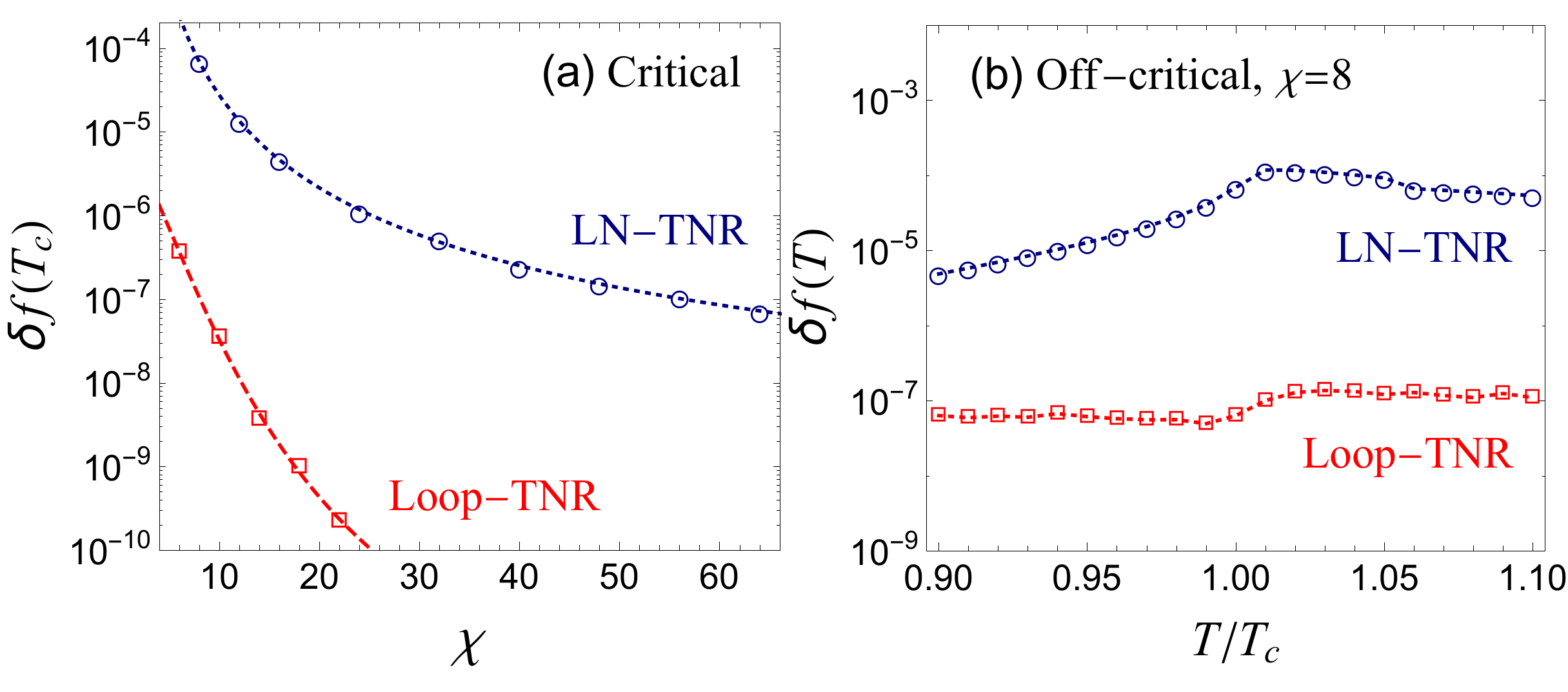}
\caption{(Color online) Comparison of the relative errors of the free energy per site computed using LN-TNR and Loop-TNR. Results were obtained on a square lattice with $2^{50}$ spins. (a) Relative error as a function of bond dimension $\chi$ at the critical point. (b) Relative error as a function of temperature for off-critical Ising models.}
\label{fig:FE}
\end{figure}

\textbf{Classical Ising model} The partition function of the 2D classical Ising model is given by $Z=\sum_{\left\{ \sigma \right\}}\exp$ $(\beta \sum_{\langle ij\rangle} \sigma_i \sigma_j)$. It can also be expressed as the contraction of a 2D tensor network with $\chi=2$ \cite{Levin:2007aa}. In this model, the spins are localised on the links of the square lattice. Each local tensor $\mathbf{T}=T^{\textrm{Ising}}_{u,l,d,r}$ has the following nonzero components:
\begin{align}
T^{\textrm{Ising}}_{1,2,1,2}&=e^{-4\beta}, &
T^{\textrm{Ising}}_{2,1,2,1}&=e^{-4\beta}, &
T^{\textrm{Ising}}_{1,1,1,1}&=e^{4\beta}, \notag \\
T^{\textrm{Ising}}_{2,2,2,2}&=e^{4\beta}, &
\textrm{others} &= 1.
\label{tnsIsing}
\end{align}

The first step is to compute the free energy of this model with $2^{50}$ spins, so that it saturates to the value of the thermodynamic limit. Fig.~\ref{fig:FE} shows the relative error of the free energy per site at and away from the critical temperature $T_{c}$. At the critical point [see Fig.~\ref{fig:FE}(a)], the error of Loop-TNR decays much faster than the error of LN-TNR. When $\chi \leq 16$, the error of Loop-TNR decays almost exponentially with $\chi$. This demonstrates a significant improvement over LN-TNR. In Fig.~\ref{fig:FE}(b), the errors of Loop-TNR remain almost constant for all temperatures near the critical point. When $\chi=8$, Loop-TNR has an accuracy in the order of $10^{-7}$. At the same point LN-TNR has an accuracy of $10^{-4} \sim 10^{-5}$. Other improved methods, such as SRG and HOSRG \cite{Xie:2009aa,Zhao:2010aa,Xie:2012aa}, can reduce the error by up to three orders of magnitude at off-critical conditions, but by only one order of magnitude at criticality. The recently proposed EV-TNR algorithm \cite{Evenbly:2015ac} can achieve the same accuracy with the same ``effective'' bond dimensions in the octagon (but a larger overall bond dimension \cite{Supplemental}). However, Loop-TNR has a lower computational cost than EV-TNR.

After applying several steps of Loop-TNR, we obtain an approximate fixed-point tensor with proper normalization and gauge fixing, which encodes the low-energy physics of the critical system. To prevent gauge fixing at the final step, $C_4$ lattice symmetry may be imposed on the RG flow. This produces a single rank-3 tensor that is approximately invariant at criticality \cite{Supplemental}.

\begin{figure}[tbp]
\includegraphics[width=1.0\columnwidth]{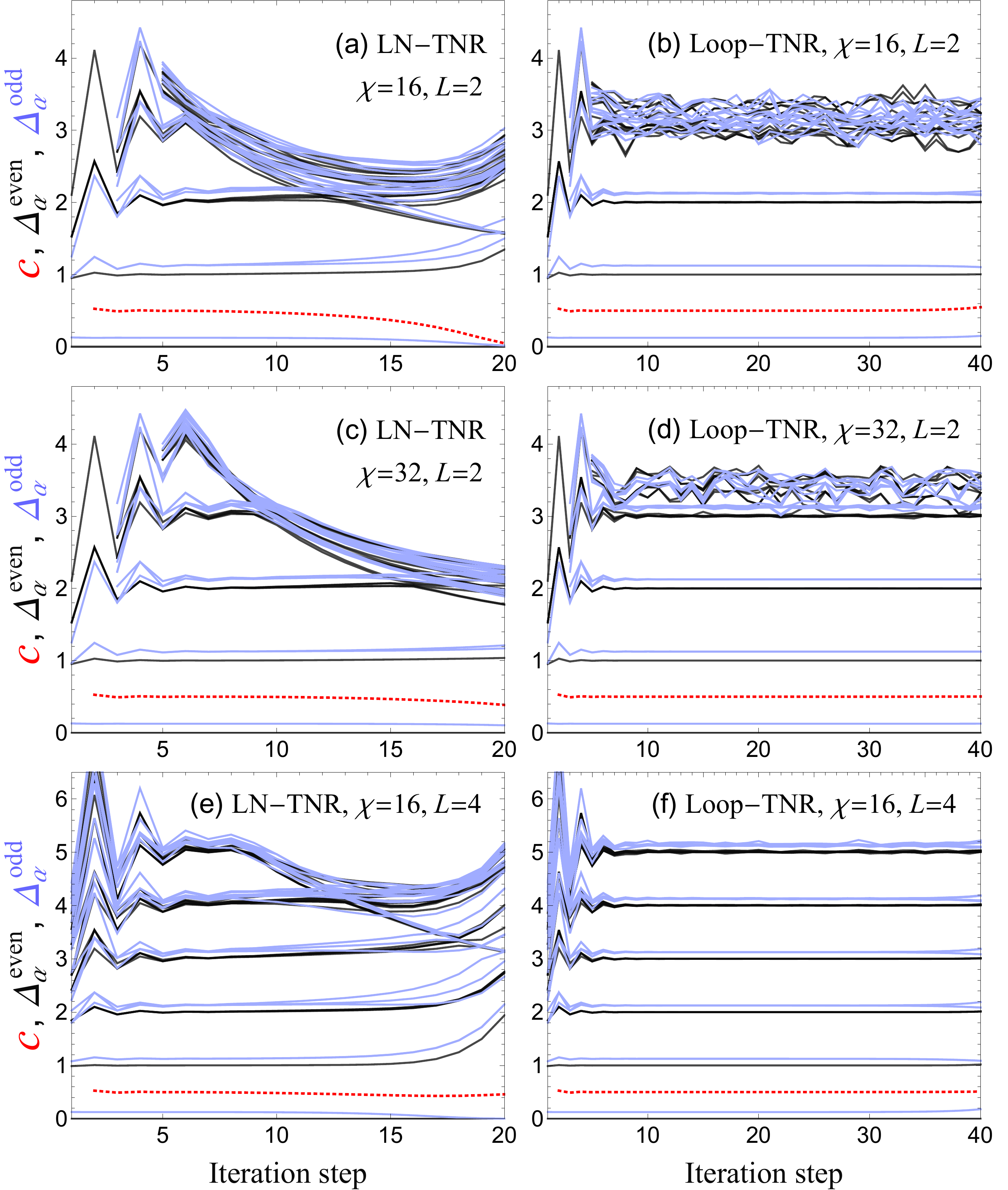}
\includegraphics[width=1.0\columnwidth]{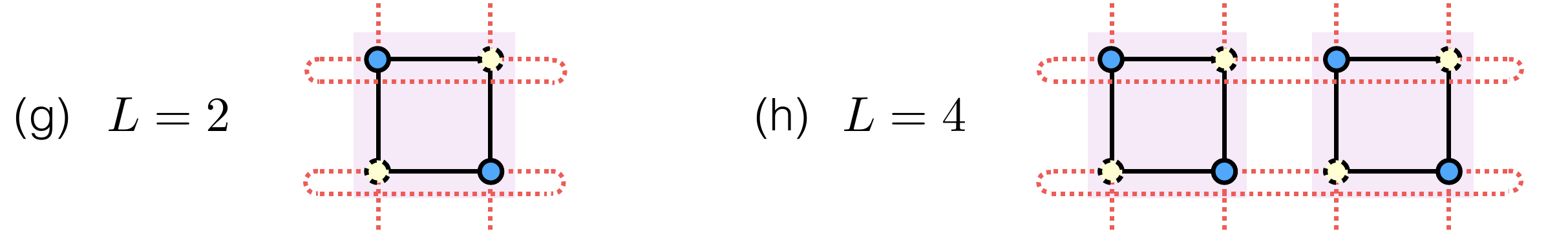}
\caption{(Color online) Comparison of central charge and scaling dimensions for LN-TNR and Loop-TNR at different iteration steps. The red dotted line denotes the central charge, the blue (light grey) solid lines denote the scaling dimensions in the $Z_2$-odd sector, and the black solid lines denote the scaling dimensions in the $Z_2$-even sector. In the $L=2$ ($L=4$) case, a transfer matrix is constructed using two (four) columns of tensors [shown in (g) and (h)]. The central charge and scaling dimensions are determined from the eigenvalues of the transfer matrix \cite{Gu:2009aa}.}
\label{fig:CH}
\end{figure}

As proposed in Ref.~\cite{Gu:2009aa}, the transfer matrix shown in Fig.~\ref{fig:CH}(g) can be constructed, and the central charge and lowest scaling dimensions determined from the eigenvalues of the transfer matrix. When $\chi=24$ and with $2^{18}$ spins, these conformal data have extremely high accuracies(up to five digits):
\begin{align*}
& c && h_1 && h_2 && h_3 
\nonumber\\
\text{Loop-TNR: } & 0.500001 && 0.1250001 && 1.000006 && 1.124994 
\nonumber\\
\text{EV-TNR: } & 0.50001 && 0.1250004 && 1.00009 && 1.12492 
\nonumber\\
\text{Exact: } & 1/2 && 1/8 && 1 && 9/8 
\end{align*}
For comparison, the central charge and the scaling dimensions obtained using EV-TNR under the same conditions are given \cite{Evenbly:2015ac} (Here $\chi$ denotes the largest bond dimension used in that scheme).

In addition to improving the accuracy of the central charge and scaling dimensions, Loop-TNR also significantly improves their stabilities. Fig.~\ref{fig:CH} compares the results from LN-TNR and Loop-TNR. In the LN-TNR case shown in the left-hand column, the high-level scaling dimensions start to merge with the low-level scaling dimensions after a few iteration steps. This indicates that the high-index ``junk'' starts to merge quickly with the low-index approximate fixed-point tensor \cite{Supplemental}. In Fig.~\ref{fig:CH}(a), the $h=2$ and $h=2.125$ scaling dimensions are destroyed by the ``junk'' after $10$ iteration steps. Correspondingly, LN-TNR fails to produce the accurate scaling dimensions, even for primary fields. In general, both stability and accuracy deteriorate at higher scaling dimensions (or, equivalently, higher-index tensor elements). 

The conformal data are significantly improved using Loop-TNR. As shown in the right-hand column of Fig.~\ref{fig:CH}, these data remain accurate up to $40$ iteration steps in the case of $\chi=16$, and even longer when $\chi=32$.  Moreover, the high-index ``junk'' is well separated from the low-index scaling dimensions. By increasing $\chi$, a greater number of scaling dimensions beyond the primary fields can be resolved from the approximate fixed-point tensors. As shown in Fig.~\ref{fig:CH} (d) and (b), the $h=3$ and $h=3.125$ scaling dimensions are clearly visible in the $\chi=32$ simulation, while they are difficult to distinguish from the high-index ``junk'' when $\chi=16$.

We have shown that for higher bond dimensions, the proper RG flow lasts longer. Thus, we believe that at the infinite $\chi$ limit, Loop-TNR can determine an \emph{infinite dimensional} fixed-point tensor described by Ising CFT at the continuum limit (with proper normalization and gauge fixing). For instance, four columns of tensors may be used to construct the transfer matrix [shown in Fig.~\ref{fig:CH}(h)], which is equivalent to using $\chi=256$. As shown in Fig.~\ref{fig:CH}(f), a greater number of scaling dimensions can be evaluated, and the accuracy is greatly improved. The result shown in Fig.~\ref{fig:CH}(f) suggests that the complete information of a CFT is encoded in the approximate fixed-point tensor.  If more tensors are used to construct the transfer matrix, it is possible to reconstruct the whole conformal tower to a given accuracy. Moreover, we have found evidence that the operator product expansion (OPE) coefficients are also encoded in the low-index approximate fixed-point tensors. How to compute these coefficients will be discussed in future work. Because the central charge, scaling dimensions, and OPE coefficients of primary fields constituent the complete set of data for a CFT, the low-index approximate fixed-point tensors can completely determine the low-energy physics with an emergent conformal symmetry. The high-index ``junk'' is subject to the conformal symmetry-breaking perturbations introduced by truncation errors, which cannot be prevented in any numerical simulations with a finite $\chi$.

\begin{figure}[tbp]
\centering
\includegraphics[width=0.85\columnwidth]{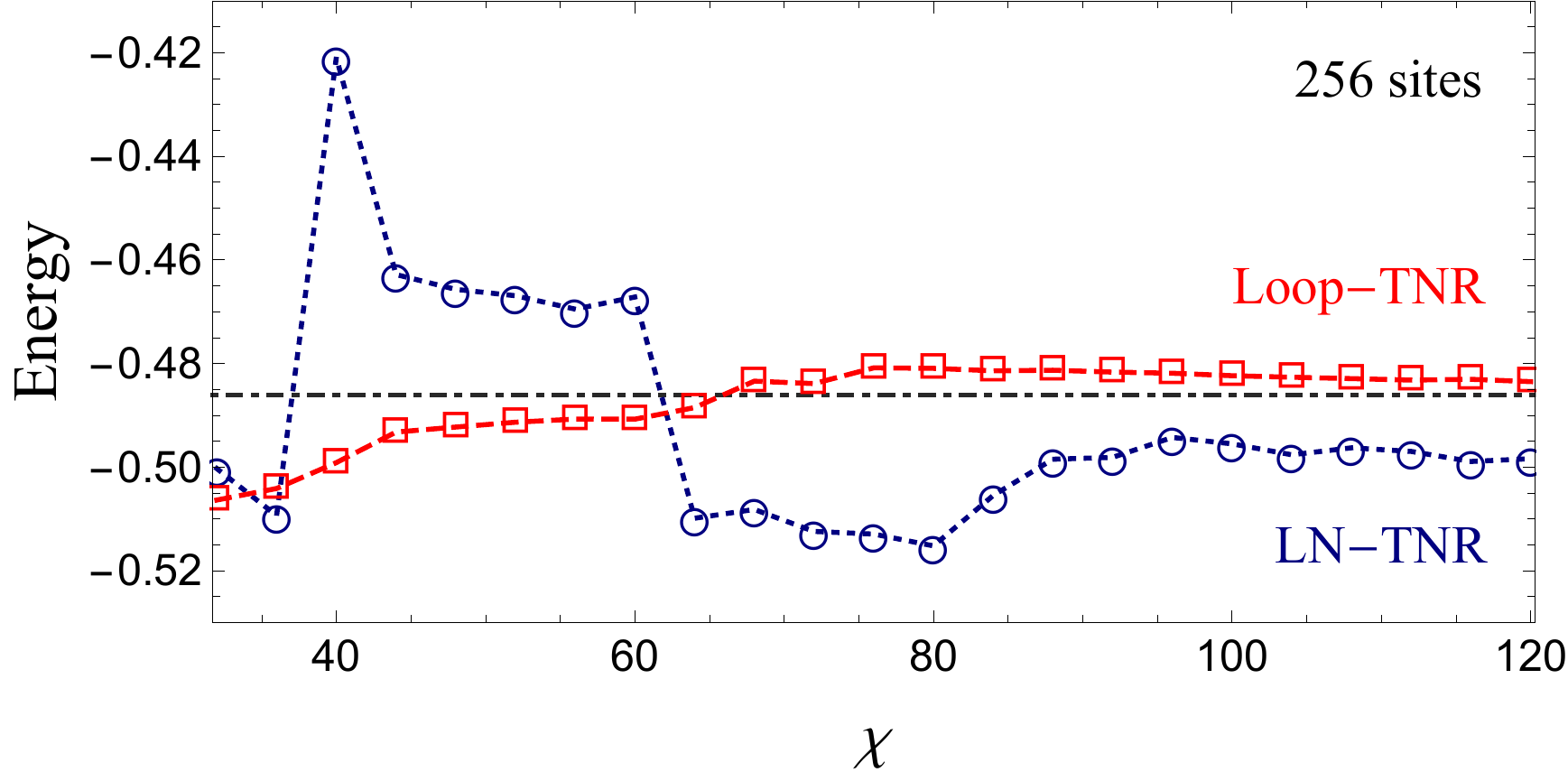}
\caption{(Color online) Benchmark of the variational energy of the $D=3$ PEPS proposed in Ref.~\cite{Wang:2013aa} for the maximally frustrated $J_1-J_2$ antiferromagnetic Heisenberg model on a square lattice (with $J_2=0.5J_1$). Here, we consider a $256$ sites system with PBC. Because the benchmark energy (dashed line) is an extrapolation for infinite systems, it could be slightly lower than the actual variational energy for $256$ sites.}
\label{fig:J1J2}
\end{figure}

\textbf{Variational energy for a 2D quantum model} Loop-TNR can compute the physical quantities of 2D projected entangled-pair states (PEPS), especially those states with divergent correlation lengths. We tested our algorithm by calculating the variational energy of the $D=3$ PEPS proposed in Ref. \cite{Wang:2013aa}. This is a variational resonating valence bond (RVB) ansatz for the $J_1-J_2$ antiferromagnetic Heisenberg model on a square lattice around the maximally frustrated regime ($J_2=0.5J_1$). The extrapolated ground state energy was obtained in Ref.~\cite{Wang:2013aa} using the boundary MPS method \cite{Jordan:2008aa,Orus:2008aa,Cirac:2011aa,Poilblanc:2012aa,Schuch:2012aa}; the value of which is shown as the black dash-dot line in Fig.~\ref{fig:J1J2}. The results of LN-TNR and Loop-TNR were calculated using a $256$-site system with periodic boundary condition (PBC). Since this PEPS has a divergent correlation length, the energy from LN-TNR is highly frustrated, and far from the accurate value. Conversely, the energy determined from Loop-TNR quickly converges to the accurate value. Here, only 20 sweeps were carried out when minimizing the cost function Eq.~(\ref{cost}) by the variational MPS method \cite{Verstraete:2004aa,Verstraete:2009aa}. Using more sweeps would have improved the results. 

\textbf{Conclusions and discussions}
We have developed the Loop-TNR algorithm, a coarse-graining transformation based on loop optimizations, to significantly improve the RG flow for both critical and off-critical systems. We demonstrated the advantage of Loop-TNR using the classical Ising model on a square lattice. High accuracy and stability of the central charge and the lowest scaling dimensions were observed at criticality. Furthermore, good accuracy was achieved in the computation of the variational energy of a frustrated 2D PEPS. 

Thanks to the concept of loop optimization, we may integrate the well-developed 1D algorithms with LN-TNR to enhance its performance. The integration with iTEBD \cite{Vidal:2007aa} gives rise to GW-TNR \cite{Gu:2009aa}, the integration with MERA \cite{Vidal:2008aa} results in EV-TNR \cite{Evenbly:2015ac}, and now the integration with variational MPS \cite{Verstraete:2004aa} leads to Loop-TNR. 
From the viewpoint of quantum field theory, our way of removing local entanglement is equivalent to \textit{integrating out} local modes during the RG transformation. As a result, Loop-TNR works better than the algorithms based on tree tensor networks (such as LN-TNR and SRG/HOSRG), where the local modes are only removed by a hard cut. For future works, we will explore the structure of the fixed-point tensor for a CFT. The 3D generalization of Loop-TNR is also a promising direction, where the ``loop-optimization'' will be replaced by the ``membrane-optimization''.

\begin{acknowledgments}
We thank Guifre Vidal and Glen Evenbly for earlier conversations about Ref.~\cite{Evenbly:2015ac} and Ref.~\cite{Gu:2009aa}. We thank the Institute for Advanced Study at Tsinghua University and the Kavli Institute for Theoretical Physics at University of California Santa Barbara for hospitality. S. Yang thanks Ling Wang and Michael Lubasch for explaining technical details in their previous publications \cite{Wang:2011ab} and \cite{Lubasch:2014aa}. 
ZCG acknowledges start-up support, Direct Grant No. 4053163 from The Chinese University of Hong Kong and the funding from RGC/ECS(No.2191110).
XGW is supported by NSF Grant No. DMR-1506475 and NSFC 11274192. He is also supported by the BMO Financial Group and the John Templeton Foundation. This research was supported in part by the National Science Foundation under Grant No. PHY11-25915. Research at Perimeter Institute is supported by the Government of Canada through Industry Canada and by the Province of Ontario through the Ministry of Economic Development \& Innovation.
\end{acknowledgments}

\bibliographystyle{apsrev4-1}
\bibliography{LoopTRG}

\onecolumngrid
\appendix
\setcounter{equation}{0}
\newpage

\renewcommand{\thesection}{S-\arabic{section}} \renewcommand{\theequation}{S%
\arabic{equation}} \setcounter{equation}{0} \renewcommand{\thefigure}{S%
\arabic{figure}} \setcounter{figure}{0}

\centerline{\textbf{Supplemental Material}}

\maketitle

\section{S-1. Detail algorithms of Loop-TNR}

\begin{figure}[tbp]
\centering
\includegraphics[width=0.5\columnwidth]{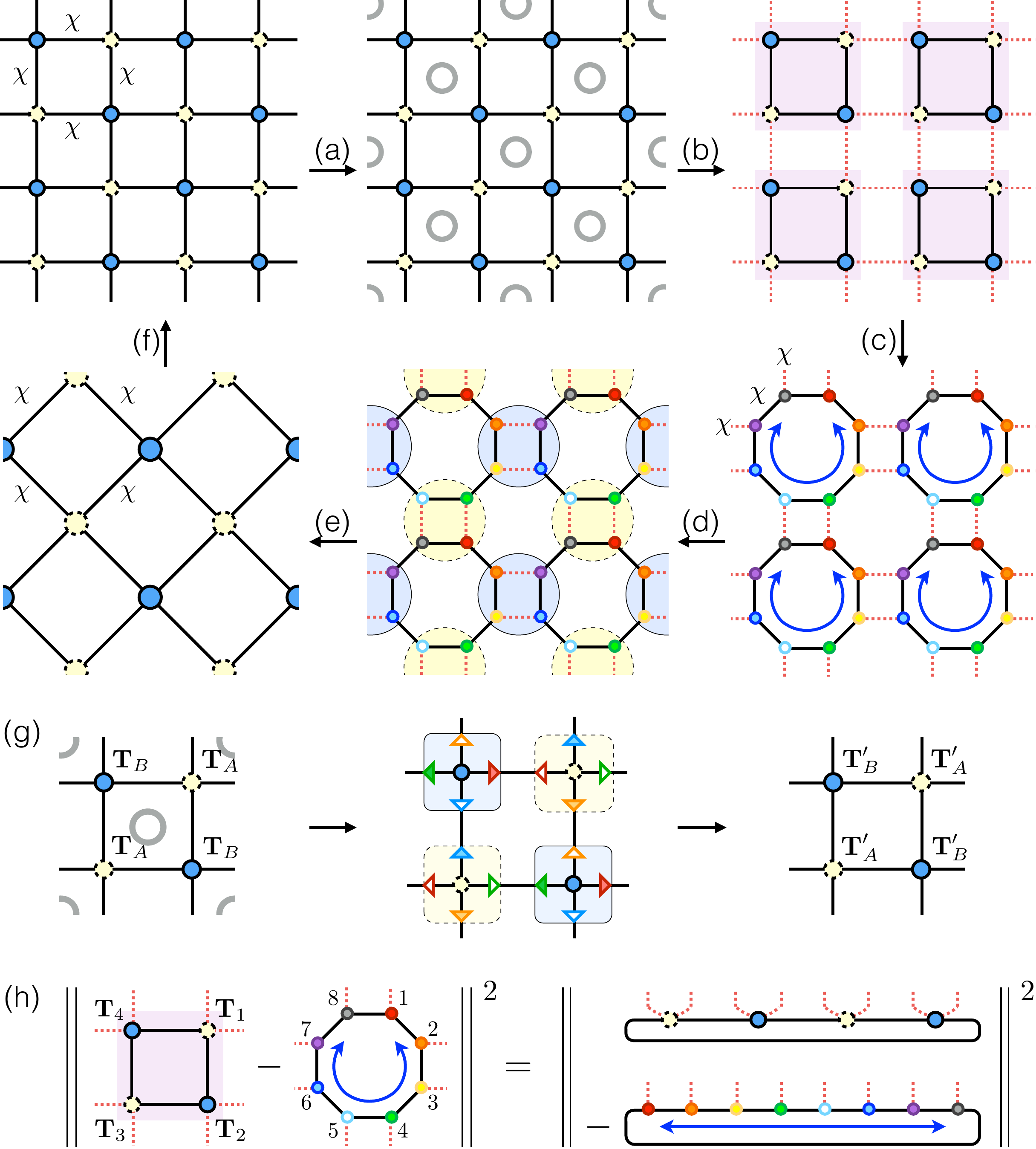}
\caption{(Color online)  
Three key steps of the Loop-TNR algorithm. (a) The entanglement filtering step. Projectors are inserted to eliminate local entanglements on the squares labeled with grey circles (see (g) for details). (c) The loop optimization step. Each of the shaded squares is deformed to a octagon made up of $8$ rank-$3$ tensors with bond dimensions no greater than $\chi$. The best approximation is found by minimizing the cost function in (h). (e) The same coarse graining step as in the standard LN-TNR algorithm. (h) The cost function of the loop optimization can be regarded as the distance between two MPS wave functions. The well-developed variational MPS method is applied to minimize the cost function.}
\label{fig:SquareNoSymS1}
\end{figure}

In this section we explain the detail algorithms of Loop-TNR. In each RG step, the loop optimization includes two parts. Part One is to filter out corner double line (CDL) tensors and to generate a canonical gauge. Part Two is to optimize each tensor so that the cost function in Fig.~\ref{fig:SquareNoSymS1}(h) is minimized. Both parts are helpful for both on- and off-critical systems. Part One and Two together can completely remove short range entanglement in each iteration step.

\subsection{A. Entanglement filtering}

The idea of entanglement filtering was first introduced in Ref.~\cite{Gu:2009aa}. Here we provide an alternative way to achieve the same goal.

\begin{figure}[tbp]
\centering
\includegraphics[width=0.75\columnwidth]{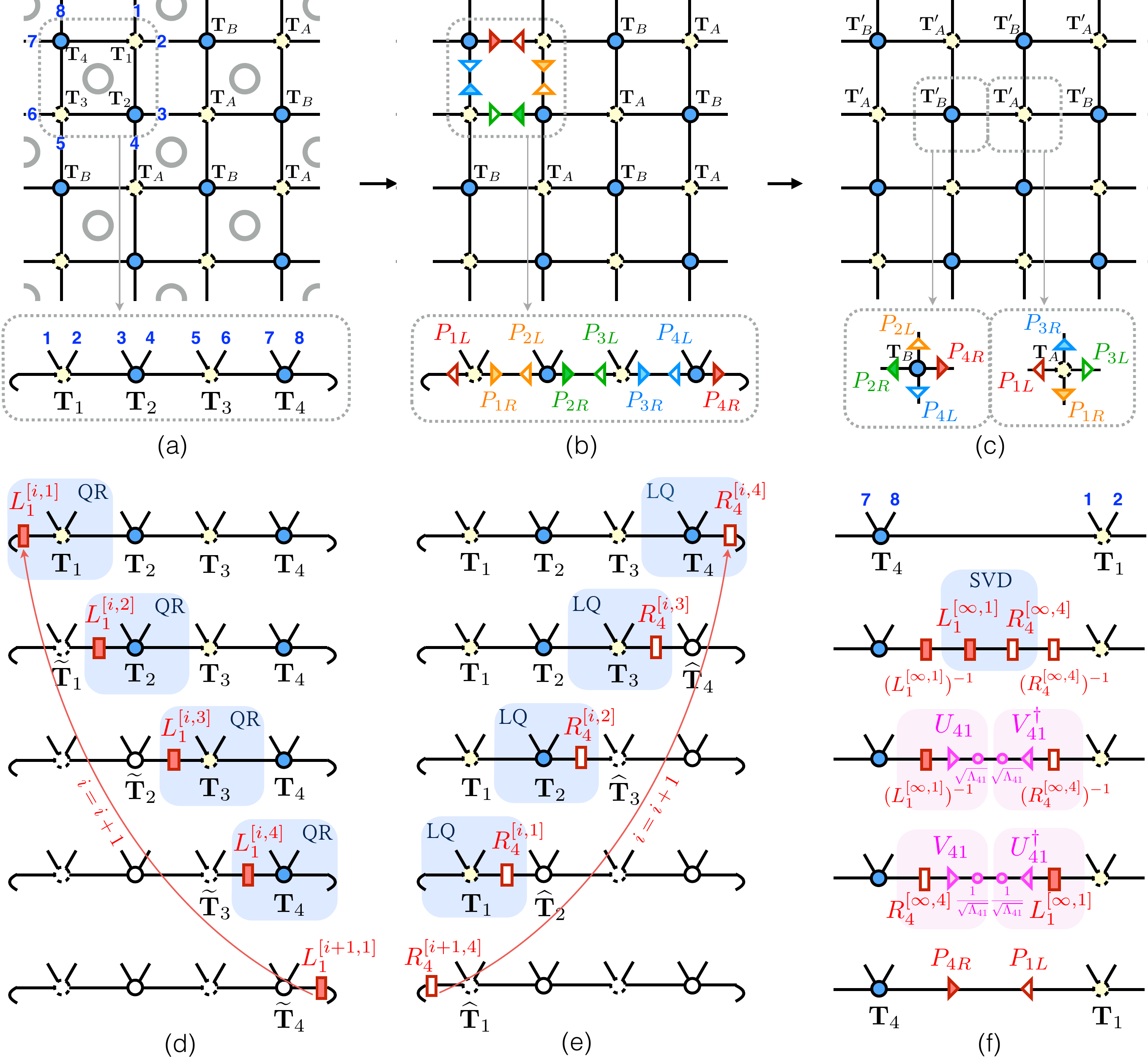}
\caption{(Color online) (a) In the dashed box, the tensors on the square can be viewed as a MPS with PBC. (b) We insert some projectors at the virtual legs of the MPS. (c) We define the new tensors by contracting the original tensors with the nearest projectors. (d) Iteratively carrying out QR decompositions to find out $\mathbf{L}_{1}^{[\infty,1]}$. (e) Iteratively carrying out LQ decompositions to find out $\mathbf{R}_{4}^{[\infty,4]}$. (f) Using $\mathbf{L}_{1}^{[\infty,1]}$ and $\mathbf{R}_{4}^{[\infty,4]}$ to find out the projectors $\mathbf{P}_{4R}$ and $\mathbf{P}_{1L}$.}
\label{fig:Loop1}
\end{figure}

We consider a square lattice with A-B sub-lattices shown in Fig.~\ref{fig:Loop1}(a). We will do filtering on every small square with a grey circle in the center. Since the system is translational invariant, we only focus on the square in the dashed box, where the tensors ($\mathbf{T}_1$, $\mathbf{T}_2$, $\mathbf{T}_3$, and $\mathbf{T}_4$) are regarded as matrix product states (MPS) on a loop. The legs across the dashed box (marked by numbers $1$-$8$) are the physical legs of the MPS, while the legs within the dashed box are the virtual legs of the MPS. With these notations in mind, our next step is to insert some projectors ($\mathbf{P}_{iL}$ and $\mathbf{P}_{iR}$ with $i=1,2,3,4$) on the virtual legs of the MPS [see Fig.~\ref{fig:Loop1}(b)]. This will not change the MPS wave function as long as all virtual legs are contracted. However, in Section S-1.C we will see it can reduce the virtual bond dimensions if the MPS is made up by double line tensors. Finally in Fig.~\ref{fig:Loop1}(c), we contract the projectors with the original tensors $\mathbf{T}_A$ and $\mathbf{T}_B$ to build new tensors $\mathbf{T}'_A$ and $\mathbf{T}'_B$ and accomplish Pare One. 

In Fig.~\ref{fig:Loop1}(d)-(f) we show how to find the projectors $\mathbf{P}_{iL}$ and $\mathbf{P}_{iR}$. We start from the $(i=1)$-th iteration in Fig.~\ref{fig:Loop1}(d), and choose $\mathbf{L}_{1}^{[i=1,1]}=\mathbb{I}$. From the first row to the second row, we group $\mathbf{L}_{1}^{[i,1]}$ and $\mathbf{T}_{1}$ and then do a QR decomposition, so that
\begin{align} 
\mathbf{L}_{1}^{[i,1]} \cdot \mathbf{T}_{1}=\widetilde{\mathbf{T}}_{1} \cdot \mathbf{L}_{1}^{[i,2]}.
\end{align}
Here $\widetilde{\mathbf{T}}_{1}$ and $\mathbf{L}_{1}^{[i,2]}$ correspond to the $\mathbf{Q}$ and $\mathbf{R}$ parts in the QR decomposition, respectively. From the second row to the third row, we do a similar QR decomposition such that
\begin{align}
\mathbf{L}_{1}^{[i,2]} \cdot \mathbf{T}_{2}=\widetilde{\mathbf{T}}_{2} \cdot \mathbf{L}_{1}^{[i,3]}.
\end{align}
We continue doing QR decompositions until we obtain $\widetilde{\mathbf{T}}_{4}$ and $\mathbf{L}_{1}^{[i+1,1]}$ in the last row of Fig.~\ref{fig:Loop1}(d). We then normalize $\mathbf{L}_{1}^{[i+1,1]}$ by a factor $\Omega_{i}$ to keep the tensor elements within the degits allowed by computers, i.e., $\mathbf{L}_{1}^{[i+1,1]}=\mathbf{L}_{1}^{[i+1,1]}/\Omega_{i}$. For example, $\Omega_{i}$ may be chosen as the largest absolute value of the element of $\mathbf{L}_{1}^{[i+1,1]}$. After that, we place $\mathbf{L}_{1}^{[i+1,1]}$ on the left side of $\mathbf{T}_{1}$ and change the iteration step from $i$ to $i+1$ (see the red arrow). We continue doing iterations until we reach a fixed point of $\mathbf{L}_{1}^{[i+1,1]}$ up to some small errors, which is further denoted as $\mathbf{L}_{1}^{[\infty,1]}$. For the Ising model it usually takes less than 20 iterations to converge. For more complicated models, we may set an upper bound of iterations. The computational cost for this step scales as $O(\chi^5)$. Similarly, in Fig.~\ref{fig:Loop1}(e) we start from $\mathbf{R}_{4}^{[i=1,4]}=\mathbb{I}$. From the first row to the second row, we group $\mathbf{T}_4$ and $\mathbf{R}_{4}^{[i,4]}$ together and then do a LQ decomposition, so that 
\begin{align}
\mathbf{T}_4 \cdot \mathbf{R}_{4}^{[i,4]} = \mathbf{R}_{4}^{[i,3]} \cdot \widehat{\mathbf{T}}_4.
\end{align}
We continue doing LQ decompositions until we reach a fixed point of the normalized $\mathbf{R}_{4}^{[i+1,4]}$, which is further denoted as $\mathbf{R}_{4}^{[\infty,4]}$. 

Once we have obtained $\mathbf{L}_{1}^{[\infty,1]}$ and $\mathbf{R}_{4}^{[\infty,4]}$, we use a similar approach introduced in Ref.~\cite{Wang:2011ab} to get the projectors $\mathbf{P}_{4R}$ and $\mathbf{P}_{1L}$. As shown in Fig.~\ref{fig:Loop1}(f), we insert an identity 
\begin{align}
\mathbb{I}=(\mathbf{L}_{1}^{[\infty,1]})^{-1} \cdot \mathbf{L}_{1}^{[\infty,1]} \cdot \mathbf{R}_{4}^{[\infty,4]} \cdot (\mathbf{R}_{4}^{[\infty,4]} )^{-1}
\end{align}
between $\mathbf{T}_4$ and $\mathbf{T}_1$. We then group $\mathbf{L}_{1}^{[\infty,1]}$ and $\mathbf{R}_{4}^{[\infty,4]}$ together and do a singular value decomposition (SVD), i.e., 
\begin{align}
\mathbf{L}_{1}^{[\infty,1]} \cdot \mathbf{R}_{4}^{[\infty,4]}= \mathbf{U}_{41} \cdot \sqrt{\mathbf{\Lambda}_{41}} \cdot \sqrt{\mathbf{\Lambda}_{41}} \cdot \mathbf{V}_{41}^{\dagger}.
\end{align}
We keep all the singular values $\mathbf{\Lambda}_{41}$ as long as they are larger than $\epsilon=10^{-12}$. Finally, the projectors $\mathbf{P}_{4R}$ and $\mathbf{P}_{1L}$ are chosen as 
\begin{align}
\mathbf{P}_{4R}=(\mathbf{L}_{1}^{[\infty,1]})^{-1} \cdot \mathbf{U}_{41} \cdot \sqrt{\mathbf{\Lambda}_{41}}, \hspace{0.5cm} \mathbf{P}_{1L}=\sqrt{\mathbf{\Lambda}_{41}} \cdot \mathbf{V}_{41}^{\dagger} \cdot (\mathbf{R}_{4}^{[\infty,4]} )^{-1}.
\label{Eq:PLPR}
\end{align}
To avoid the matrix inversion, we use the fact that
\begin{align}
(\mathbf{L}_{1}^{[\infty,1]} \mathbf{R}_{4}^{[\infty,4]})^{-1}=(\mathbf{R}_{4}^{[\infty,4]} )^{-1} (\mathbf{L}_{1}^{[\infty,1]})^{-1}= \mathbf{V}_{41} \frac{1}{\mathbf{\Lambda}_{41}} \mathbf{U}_{41}^{\dagger}.
\end{align}
Therefore,
\begin{align}
(\mathbf{L}_{1}^{[\infty,1]})^{-1}=\mathbf{R}_{4}^{[\infty,4]} \mathbf{V}_{41} \frac{1}{\mathbf{\Lambda}_{41}} \mathbf{U}_{41}^{\dagger}, \hspace{0.5cm} (\mathbf{R}_{4}^{[\infty,4]})^{-1}=\mathbf{V}_{41} \frac{1}{\mathbf{\Lambda}_{41}} \mathbf{U}_{41}^{\dagger} \mathbf{L}_{1}^{[\infty,1]}.
\label{Eq:LinvRinv}
\end{align}
Inserting Eq.~(\ref{Eq:LinvRinv}) into Eq.~(\ref{Eq:PLPR}) we finally have
\begin{align}
\mathbf{P}_{4R}=\mathbf{R}_{4}^{[\infty,4]} \mathbf{V}_{41} \frac{1}{\sqrt{\mathbf{\Lambda}_{41}}}, \hspace{0.5cm} \mathbf{P}_{1L}=\frac{1}{\sqrt{\mathbf{\Lambda}_{41}}} \mathbf{U}_{41}^{\dagger} \mathbf{L}_{1}^{[\infty,1]}.
\end{align}
We then use the same method to obtain all the projectors $\mathbf{P}_{iL}$ and $\mathbf{P}_{iR}$ and insert them at the positions shown in Fig.~\ref{fig:Loop1}(b). Please note that we do not need to make any approximation in Part One. Thus the partition function of the 2D classical system remains unchanged after we tracing out all inner indices. In fact, what we have changed is only the local gauge of each tensor, which is helpful for producing the correct fixed points. However, if $\mathbf{T}_{A}$ and $\mathbf{T}_{B}$ is made up by CDL tensors, our approach can reduce the number of non-zero singular values in $\mathbf{\Lambda}$, and thus reduce the bond dimensions of the new tensors $\mathbf{T}'_{A}$ and $\mathbf{T}'_{B}$. In Section S-1.C we will explain the reason for that.

To sum up, this part provides a local canonical gauge and can filter out CDL tensors. It is crucial for generating the correct fixed point at off-critical points. It is also helpful for enhancing the performance in Part Two. The computational cost of this part scales as $O(\chi^5)$. 

\subsection{B. Optimizing tensors on a loop}

\begin{figure}[tbp]
\centering
\includegraphics[width=0.8\columnwidth]{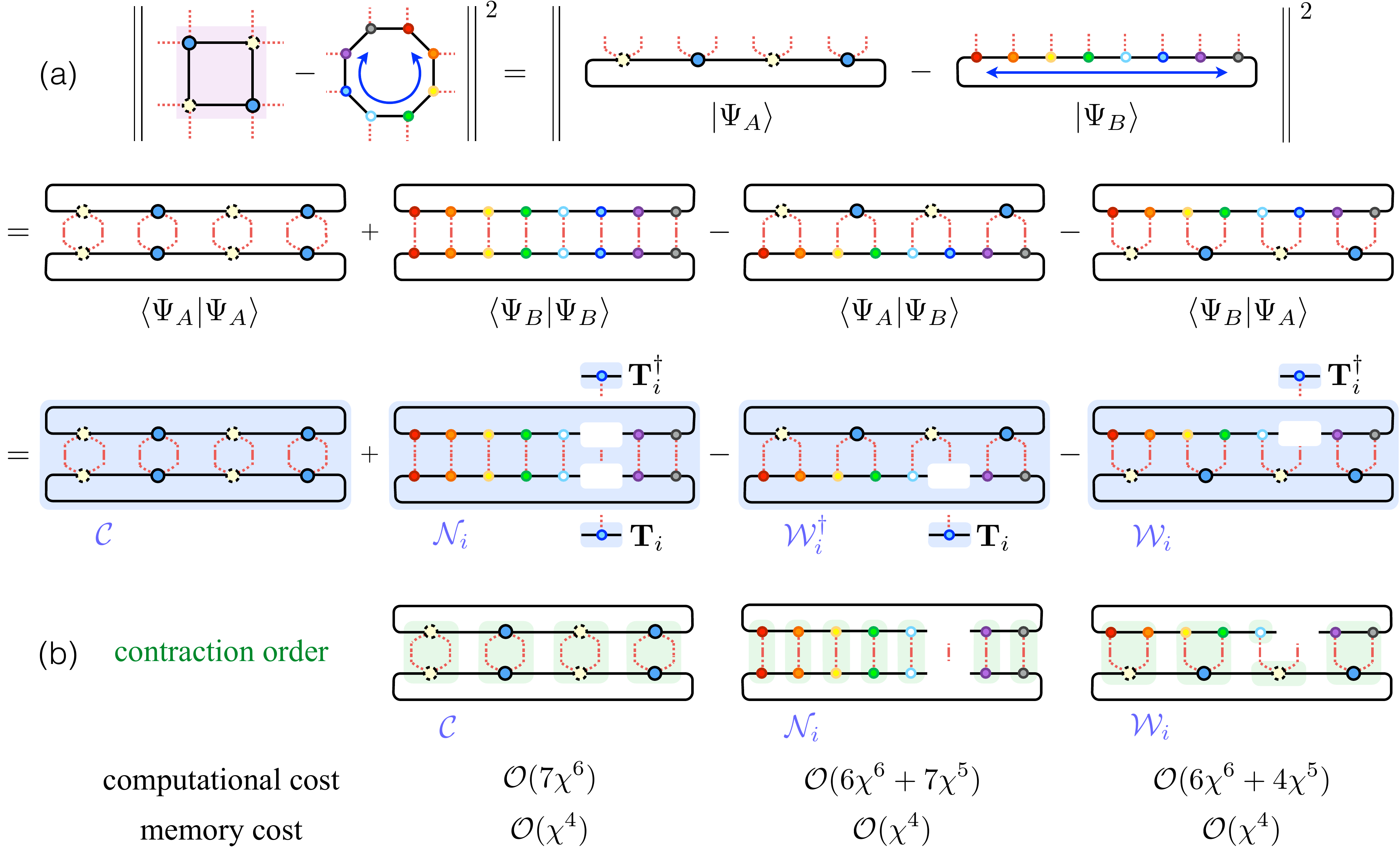}
\caption{(Color online) (a) The cost function is a quadratic function of the local tensor $\mathbf{T}_{i}$. (b) The contraction order and the computational cost for loop optimization.}
\label{fig:Loop3}
\end{figure}

In this part we will find the optimized tensors which can minimize the cost function in Fig.~\ref{fig:SquareNoSymS1}(g) [see also Fig.~\ref{fig:Loop3}]. Since the tensors on a loop can be regarded as a $8$-site MPS with periodic boundary condition, we may use the standard variational MPS method \cite{Verstraete:2004aa,Verstraete:2009aa} to solve the cost function. 

As shown in Fig.~\ref{fig:Loop3}, the the cost function is a quadratic function for parameters associated with one tensor $\mathbf{T}_{i}$, 
\begin{align}
f(\mathbf{T}_{i}) &=\left\| \left| \Psi_{A} \right\rangle - \left| \Psi_{B} \right\rangle \right\|= \left\langle \Psi_{A} | \Psi_{A} \right\rangle+\left\langle \Psi_{B} | \Psi_{B} \right\rangle -\left\langle \Psi_{A} | \Psi_{B} \right\rangle -\left\langle \Psi_{B} | \Psi_{A} \right\rangle \notag \\
&= \mathcal{C} + \mathbf{T}_{i}^{\dagger} \mathcal{N} _{i} \mathbf{T}_{i}- \mathcal{W}_{i}^{\dagger} \mathbf{T}_{i} -\mathbf{T}_{i}^{\dagger} \mathcal{W}_{i},
\end{align}
Suppose all other tensors are fixed except $\mathbf{T}_{i}$, the minimum of $f(\mathbf{T}_{i})$ can be found by solving the linear equation
\begin{align}
\mathcal{N}_{i} \mathbf{T}_{i}=\mathcal{W}_{i}.
\end{align}
After optimizing $\mathbf{T}_{i}$, we move to the next site and do this kind of optimization site-by-site. We sweep back and forth until $f$ converges to a small value. Here $\mathcal{N}_{i}$ and $\mathcal{W}_{i}$ are calculated in an efficient way by carefully choosing the order of tensor contractions. More concretely, we first contract the tensors within each shaded block in Fig.~\ref{fig:Loop3}, and then contract among the shaded blocks. Therefore, the overall computational cost of this part scales as $O(\chi^6)$. 

As to the initial states, we may simply choose the tensors from LN-TNR. Alternatively, we may first roughly truncate the bonds by considering the environments on a loop, and then optimize tensors site by site. We find a good starting point can greatly speed up the convergence \cite{Orus:2008aa,Wang:2011ab}.

We emphasize that our algorithm is very general. We do not need to impose any unitary condition as required for EV-TNR \cite{Evenbly:2015ac,Evenbly:2017aa}. It works equally well for calculating the partition functions of 2D classical systems, Euclidean path integrals of $(1+1)$D quantum systems, and physical quantities of 2D quantum systems. 

\subsection{C. Filtering out double conner line tensors}

In this subsection we show the method introduced in Section S-1.A can filter out CDL tensors.

\begin{figure}[tbp]
\centering
\includegraphics[width=0.75\columnwidth]{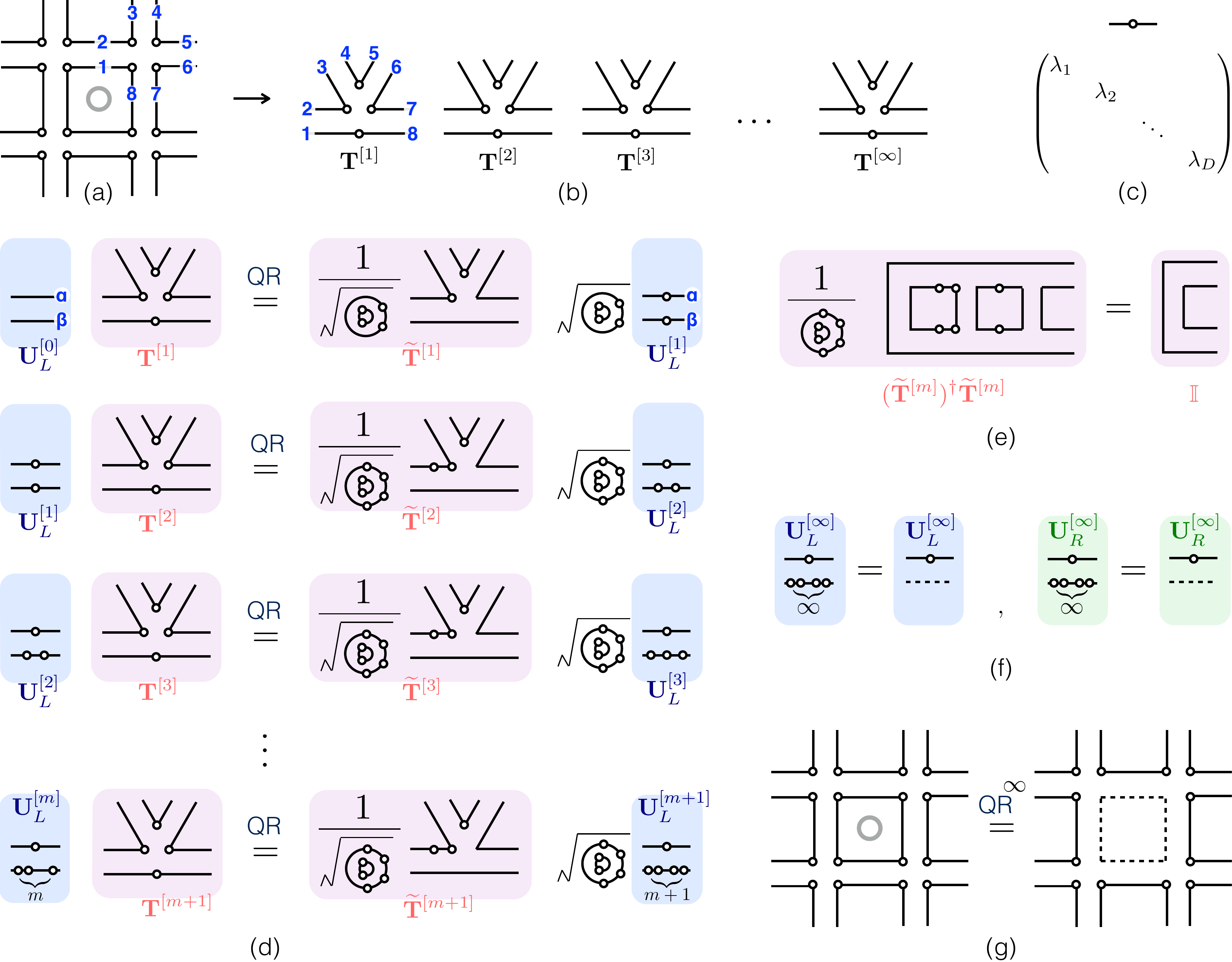}
\caption{(Color online) (a) CDL tensors on a square lattice. (b) We view the CDL tensors in (a) as a MPS and repeat the unit cell for infinite times. (c) A solid line with an empty dot on it represents a diagonal matrix. (d) Iteratively carrying out QR decompositions to find out $\mathbf{U}_{L}^{[m+1]}$. (e) $\widetilde{\mathbf{T}}^{[m]}$ satisfies the canonical condition. (f) After many iterations, $\mathbf{U}_{L}^{[\infty]}$ and $\mathbf{U}_{R}^{[\infty]}$ can be simplified to a single-line matrix. (g) The inner circle can be filtered out using the entanglement filtering approach.}
\label{fig:CDL}
\end{figure}

We consider a translational invariant tensor network on a square lattice shown in Fig.~\ref{fig:CDL}(a). Each local tensor $\mathbf{T}$ has a CDL structure,
\begin{align}
\mathbf{T}=T_{u;l;d;r}=T_{(j_3,j_4);(j_2,j_1);(j_8,j_7);(j_5,j_6)}=\mathbf{\Lambda}_{32} \mathbf{\Lambda}_{18} \mathbf{\Lambda}_{76} \mathbf{\Lambda}_{54} \delta_{j_3,j_2} \delta_{j_1,j_8} \delta_{j_7,j_6} \delta_{j_5,j_4}.
\label{Eq:CDL-T}
\end{align}
For simplicity, hereafter we choose 
\begin{align}
\mathbf{\Lambda}_{32}= \mathbf{\Lambda}_{18}= \mathbf{\Lambda}_{76} =\mathbf{\Lambda}_{54}=\mathbf{\Lambda}=\mathrm{diag}(\lambda_{1}, \lambda_{2}, \cdots, \lambda_{D}),
\end{align}
where $\lambda_{1} > \lambda_{2} > \cdots > \lambda_{D}$ (see Fig.~\ref{fig:CDL}(c)). However, our derivations below may be easily generalized to other cases. Graphically, $\mathbf{\Lambda}$ is denoted as an empty dot, while the $\delta$-functions in Eq.~(\ref{Eq:CDL-T}) are denoted as solid lines. 

In Fig.~\ref{fig:CDL}(a), the four tensors around the grey circle may be regarded as a four-site MPS wave function with periodic boundary condition. We may write the CDL tensors in MPS forms in Fig.~\ref{fig:CDL}(b), where they becomes $\mathbf{T}^{[1]}$ to  $\mathbf{T}^{[4]}$. We repeat the 4-site unit cell for infinite times. Therefore the end of the MPS chain is labeled as $\mathbf{T}^{[\infty]}$. Now we use the approach in Fig.~\ref{fig:Loop1} to filter out the CDL tensor. On the left side of $\mathbf{T}^{[1]}$ we insert an identity matrix $\mathbf{U}_{L}^{[0]}$, which is denoted as a double line without any empty dot. We group $\mathbf{U}_{L}^{[0]}$ and $\mathbf{T}^{[1]}$ together and do a QR decomposition shown in Fig.~\ref{fig:CDL}(d). The Q part becomes $\widetilde{\mathbf{T}}^{[1]}$, while the R part becomes $\mathbf{U}_{L}^{[1]}$ multiplied by a normalization factor. The normalization factor is carefully chosen to make sure $\widetilde{\mathbf{T}}^{[m]}$ satisfies the unitary condition in Fig.~\ref{fig:CDL}(e). We note that this QR decomposition actually moves $\mathbf{\Lambda}_{76}$ from $\mathbf{T}^{[1]}$ to the $\alpha$-leg of $\mathbf{U}_{L}^{[1]}$. At the same time, it moves $\mathbf{\Lambda}_{18}$ from $\mathbf{T}^{[1]}$ to the $\beta$-leg of $\mathbf{U}_{L}^{[1]}$. On the second line of Fig.~\ref{fig:CDL}(d), $\mathbf{U}_{L}^{[1]}$ and $\mathbf{T}^{[2]}$ are decomposed into $\widetilde{\mathbf{T}}^{[2]}$ and $\mathbf{U}_{L}^{[2]}$ using QR decomposition. We note that this actually moves $\mathbf{\Lambda}_{76}$ from the $\alpha$-leg of $\mathbf{U}_{L}^{[1]}$ to the $2$-leg of $\widetilde{\mathbf{T}}^{[2]}$. It also moves  $\mathbf{\Lambda}_{76}$ from $\mathbf{T}^{[2]}$ to the $\alpha$-leg of $\mathbf{U}_{L}^{[2]}$, and moves $\mathbf{\Lambda}_{18}$ from both $\mathbf{U}_{L}^{[1]}$ and $\mathbf{T}^{[2]}$ to the $\beta$-leg of $\mathbf{U}_{L}^{[2]}$. Therefore, after this step there are two empty dots on the $\beta$-leg of $\mathbf{U}_{L}^{[2]}$. We continue doing QR decompositions as indicated in Fig.~\ref{fig:CDL}(d). We see that the number of empty dots are accumulating on the $\beta$-leg of $\mathbf{U}_{L}^{[m]}$. In particular, after $m$ times of QR decomposition, there will be one empty dot on the $\alpha$-leg of $\mathbf{U}_{L}^{[m]}$, and $m$ empty dots on the $\beta$-leg of $\mathbf{U}_{L}^{[m]}$. This means $\mathbf{U}_{L}^{[m]}=\mathbf{\Lambda}_{76} \otimes (\mathbf{\Lambda}_{18})^{m}$. When $m \rightarrow \infty$, only the dominant element in $\mathbf{\Lambda}_{18}$ left, $\mathbf{U}_{L}^{[m]}=\mathbf{\Lambda}_{76} \times (\lambda_{1})^{m}$. Therefore the $\beta$-leg of $\mathbf{U}_{L}^{[m]}=$ is reduced to dimension one, which is denoted as the dashed line in Fig.~\ref{fig:CDL}(f). We may get the same conclusion using LQ decompositions. After that, we use the approach introduced in Fig.~\ref{fig:Loop1}(f) to find the projectors and apply them to the CDL tensor. This results in a dashed line square in Fig.~\ref{fig:CDL}(g). We do the same thing on every square with a grey circle in the middle. Finally, after one step of renormalization we can filter out all the CDL tensors. 

\subsection{D. Constructing the transfer matrix}

After each coarse-graining step shown in Fig.~\ref{fig:SquareNoSymS1}(e), we construct the transfer matrix according to Fig. 3(g) or Fig. 3(h) in the main text. Please note that the variational MPS method used in Section S-1.B may introduce an arbitrary gauge choice for every local tensor. On each virtual bond one may insert an arbitrary $u$ and $u^{-1}$ that does not change the MPS wave function, and also does not change the whole partition function. We would like to point out, this arbitrary gauge choice only transforms the transfer matrix from $\mathcal{M}$ to $U \mathcal{M} U^{-1}$, which will not alter its eigenvalues. Therefore all physical quantities obtained from the transfer matrix, including the central charge and scaling dimensions, remain the same. 

\section{S-2. Comparisons of central charge and scaling dimensions}

\begin{figure}[tbp]
\centering
\includegraphics[width=0.4\columnwidth]{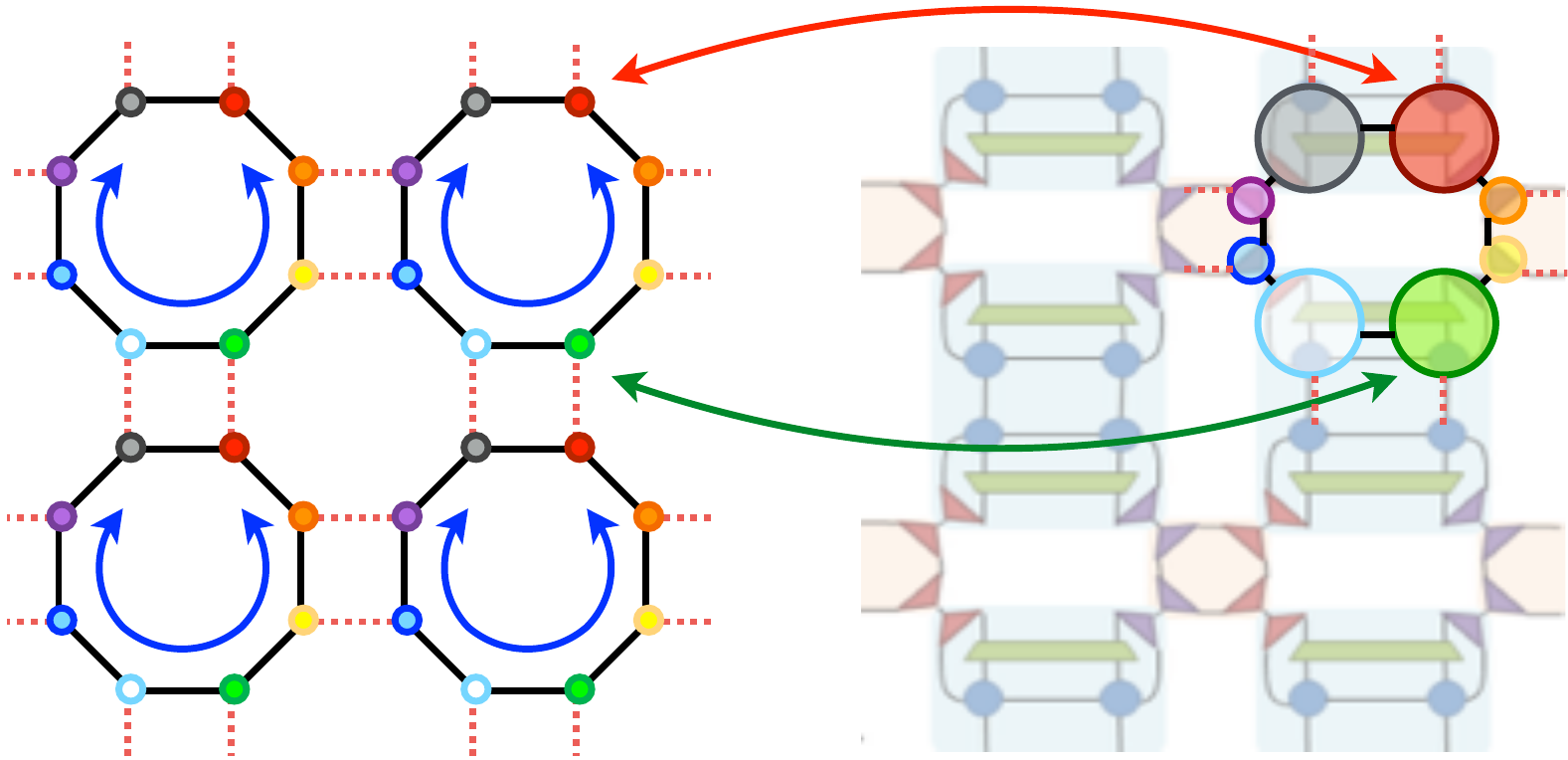}
\caption{(Color online) Relations between Loop-TNR (left) and EV-TNR \cite{Evenbly:2015ac} (right).}
\label{fig:Relation}
\end{figure}

In this section, we compare the central charge and the lowest scaling dimensions obtained from different approaches. 

First of all, all TNR algorithms can produce quite accurate central charge and scaling dimensions. However, both Loop-TNR and EV-TNR are more accurate than LN-TNR. 

Moreover, when using the $L=2$ transfer matrix, the accuracy of Loop-TNR with $\chi=16$ is comparable with the accuracy of EV-TNR with $\chi=24$. When using both $\chi=24$ and $L=2$, the accuracy of Loop-TNR is higher than the one of EV-TNR. Here $\chi$ denotes the largest bond dimension used in EV-TNR, although the authors actually use different bond dimensions at different steps \cite{Evenbly:2015ac}. Since EV-TNR can be regarded as a special case of Loop-TNR [see Fig.~\ref{fig:Relation}], the $\chi=24$ case in EV-TNR actually corresponds to the $\chi=16$ case in Loop-TNR when optimizing the cost functions on a loop. That is probably why their results are comparable. However, the overall computational cost scales in the order of the largest bond dimension $\chi$. (For Loop-TNR it is $\mathcal{O}(\chi^6)$ and for EV-TNR it is $\mathcal{O}(\chi^7)$ or $\mathcal{O}(\chi^6)$ \cite{Evenbly:2015ac,Evenbly:2017aa} \cite{Note1}). Therefore we still need to compare the results under the same largest bond dimension, then Loop-TNR seems better. 

Furthermore, the results are more accurate using a larger bond dimension $\chi$. They are getting better using a larger transfer matrix, i.e., the $L=2$ LN-TNR is better than the $L=1$ LN-TNR, and the $L=4$ cases are generally better than the corresponding $L=2$ cases. In fact, we may regard the effective bond dimension of the transfer matrix as $\chi^{L}$, the larger the better.

\begin{center}
\begin{tabular}{cc|ccccccc}
\hline 
 & Exact & LN-TNR & LN-TNR & Loop-TNR & Loop-TNR & Loop-TNR & Loop-TNR & EV-TNR \cite{Evenbly:2015ac} \tabularnewline
 &  & $\chi=64$ & $\chi=64$ & $\chi=16$ & $\chi=24$ & $\chi=16$ & $\chi=24$ & $\chi=24$\tabularnewline
 &  & $L=1$ & $L=2$ & $L=2$ & $L=2$ & $L=4$ & $L=4$ & $L=2$\tabularnewline
 &  & $2^{11}$ spins & $2^{11}$ spins & $2^{18}$ spins & $2^{18}$ spins & $2^{18}$ spins & $2^{18}$ spins & $2^{18}$ spins\tabularnewline
\hline 
$c$ & 0.5 & 0.49946958 & 0.49970058 & 0.50001491 & 0.50000165 & 0.50009255 & 0.50008794 & 0.50001\tabularnewline
\hline 
$\sigma$ & 0.125 & 0.12504027 & 0.12500837 & 0.12500528 & 0.12500011 & 0.12501117 & 0.12499789 & 0.1250004\tabularnewline
$\epsilon$ & 1 & 1.00028269 & 0.99996784 & 1.00000566 & 1.00000601 & 0.99999403 & 1.00000507 & 1.00009\tabularnewline
 & 1.125 & 1.12368834 & 1.12444247 & 1.12495187 & 1.12499400 & 1.12498755 & 1.12500559 & 1.12492\tabularnewline
 & 1.125 & 1.12394625 & 1.12450246 & 1.12510600 & 1.12500464 & 1.12498755 & 1.12500559 & 1.12510\tabularnewline
 & 2 & 1.92334948 & 1.99811859 & 2.00000743 & 1.99970911 & 1.99999517 & 2.00000985 & 1.99922\tabularnewline
 & 2 & 1.96264143 & 1.99815644 & 2.00066117 & 2.00016629 & 1.99999517 & 2.00000985 & 1.99986\tabularnewline
 & 2 & 1.97496787 & 1.99868822 & 2.00066117 & 2.00031103 & 2.00002744 & 2.00001690 & 2.00006\tabularnewline
 & 2 & 2.00274974 & 1.99948966 & 2.00586886 & 2.00131384 & 2.00006203 & 2.00002745 & 2.00168\tabularnewline
\hline 
\end{tabular}
\end{center}

\section{S-3. Loop-TNR with lattice symmetries}

\subsection{A. Detailed algorithms}

\begin{figure}[tbp]
\centering
\includegraphics[width=0.6\columnwidth]{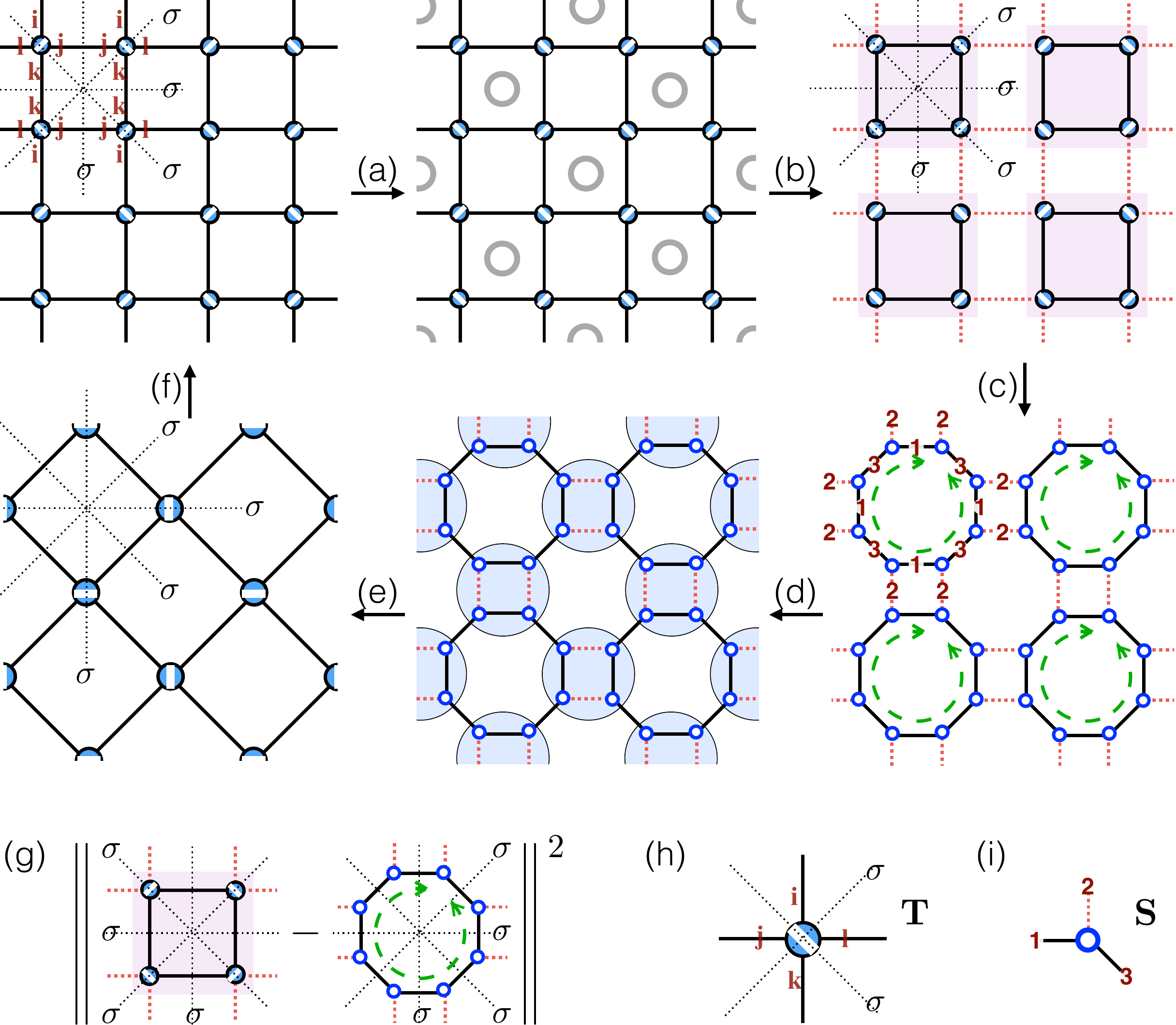}
\caption{(Color online) (a)-(f) Coarse-graining transformation of the Loop-TNR with lattice symmetries, where $\sigma$ denotes the axis of symmetry. (g) Cost function of the loop optimization. (h) Building block of the square lattice. (i) Building block of the square-octagon lattice.}
\label{fig:SquareSym}
\end{figure}

For the Loop-TNR method introduced in the main text, the tensors $\mathbf{S}_1,\mathbf{S}_2, \cdots \mathbf{S}_8$ in Fig.~\ref{fig:SquareNoSymS1}(h) are all different from each other. However, for highly isometric tensor networks, we may use an alternative approach to minimize the cost function shown in
Fig.~\ref{fig:SquareSym}(g). So that the $\mathbf{S}$ tensors are related to each other by reflection or translation symmetries. Furthermore, with a special arrangement of local tensors, we can keep the global $C_4$ and reflection symmetries of the tensor network. 

We illustrate the renormalization transformation in Fig.~\ref{fig:SquareSym}. For the classical Ising model on a square lattice, the initial $\chi=2$ tensor $\mathbf{T}=T_{ijkl}$ has the following symmetries [see Fig.~\ref{fig:SquareSym}(h)].
\begin{align}
T_{ijkl}=T_{ilkj}=T_{kjil}=T_{jilk}=T_{lkji}.
\end{align}
This allows us to enlarge the unit cell and relabel the square tensor network. As shown in Fig.~\ref{fig:SquareSym}(a), each unit cell contains four tensors, and each tensor is related to the nearest neighbours by reflection symmetries (denoted by $\sigma$). After this re-labeling, we only need the local tensor $T_{ijkl}$ satisfies
\begin{align}
T_{ijkl}=T_{jilk}=T_{lkji}.
\end{align}
With this, we find the initial tensor network has a global $C_4$ symmetry with respect to the center of the unit cell. 

It turns out that we may keep all the above symmetries in the entanglement filtering step. Next, when deforming the square lattice to a square-octagon lattice, we are aiming to find a highly symmetric octagon network that minimize the cost function Fig.~\ref{fig:SquareSym}(g). Here, each octagon [see Fig.~\ref{fig:SquareSym}(d)] is made up by a single rank-$3$ tensor $\mathbf{S}$ by reflection symmetries. The building block $\mathbf{S}=S_{i_1,i_2,i_3}$ is shown in Fig.~\ref{fig:SquareSym}(i). We then use the conjugate gradient method introduced in Ref.~\cite{Pirvu:2011aa,Pirvu:2012aa} to find the optimal $\mathbf{S}$. Finally in Fig.~\ref{fig:SquareSym}(e), we contract over the inner indicies within the circles and obtain the new square lattice. Since the octagon network has four axes of symmetry and the square network in the blue circle has two axes of symmetry, the coarse-grained tensor network on the square lattice shares the same symmetries of the original tensor network. i.e., each unit cell has four axis of symmetry and a global $C_4$ symmetry with respect to the center of the unit cell.

By keeping lattice symmetries in each iteration step, we have automatically fixed the gauge of the building block $\mathbf{S}$. In the ideal case with infinite bond dimensions, we will end up with an absolutely invariant fixed point tensor $\mathbf{S}$ using this kind of Loop-TNR. However, there is no absolutely invariant fixed point tensor with finite bond dimensions. In practice, we find the individual tensor elements obtained by Loop-TNR are still slightly different from step to step. In the following we compare the tensor elements obtained from LN-TNR and Loop-TNR.

\subsection{B. Fixed point tensors}

\begin{figure}[tbp]
\centering
(a)\includegraphics[width=1.0\columnwidth]{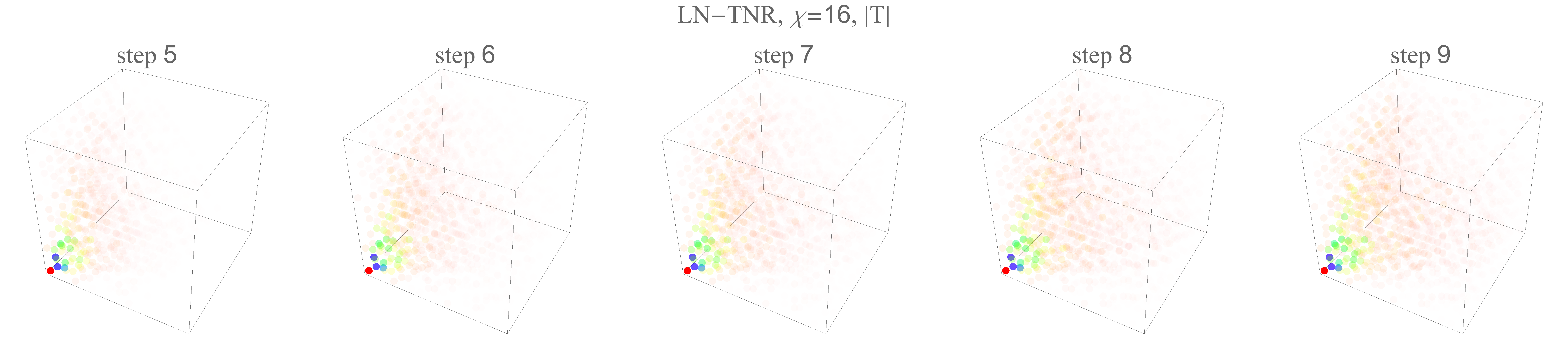}
(b)\includegraphics[width=0.8\columnwidth]{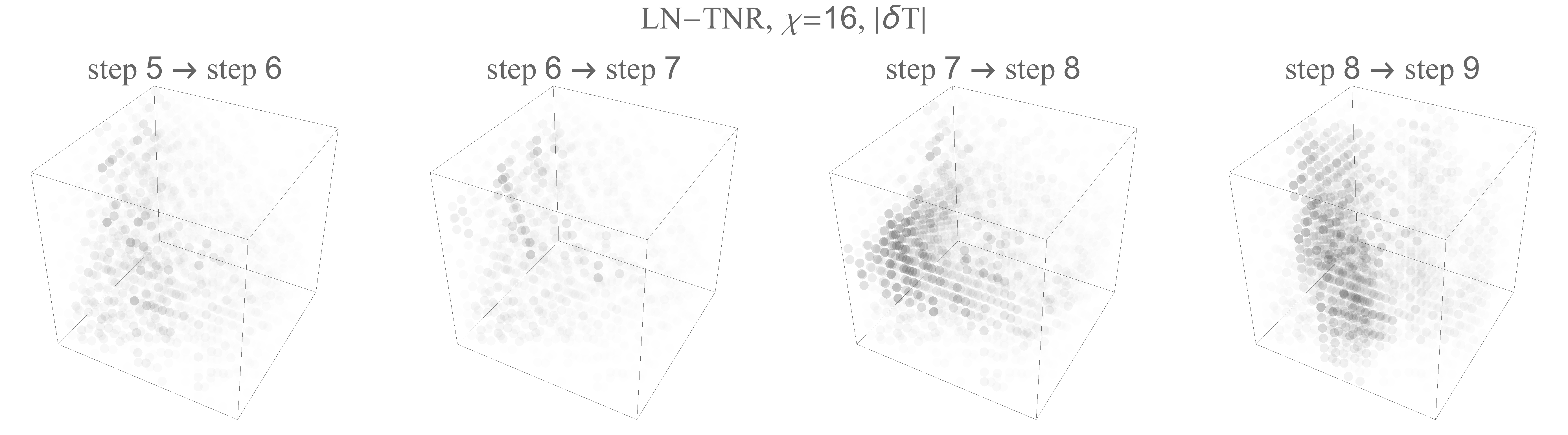}
(c)\includegraphics[width=1.0\columnwidth]{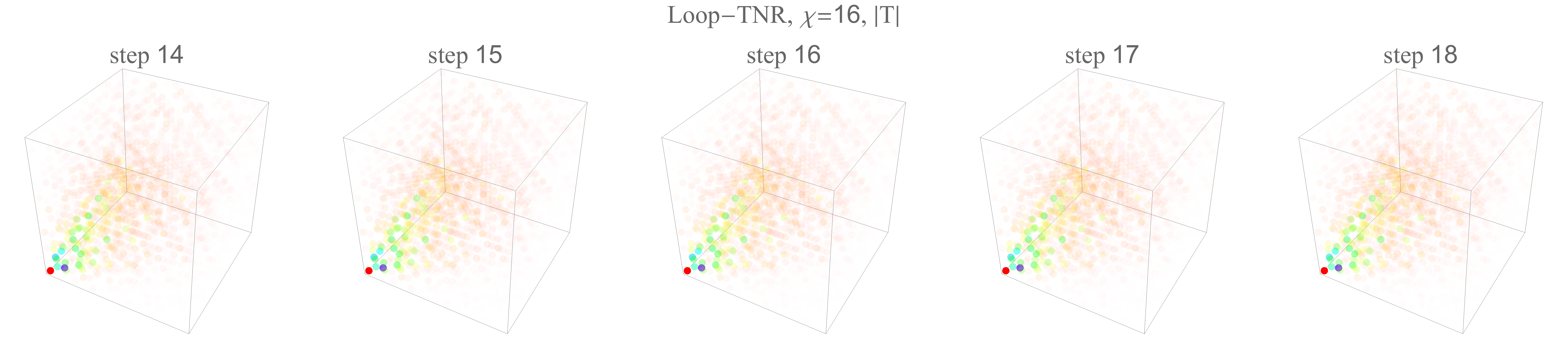}
(d)\includegraphics[width=0.8\columnwidth]{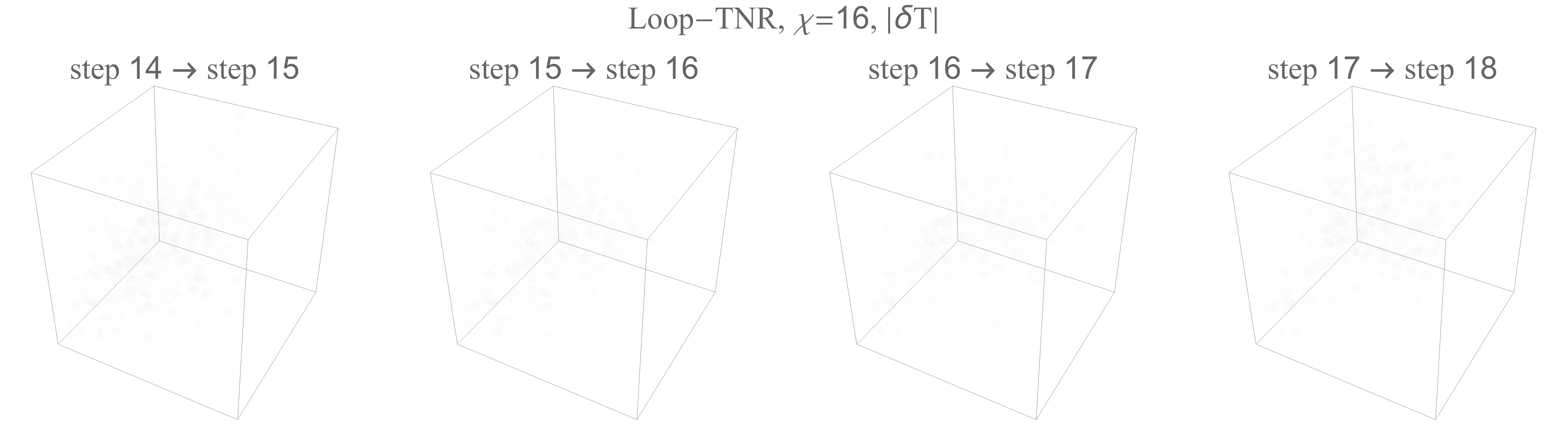}
\caption{(Color online) (a) and (c): The absolute values of individual tensor elements obtained using LN-TNR (a) and Loop-TNR (c). Color and opacity denote the amplitude. (b) and (d): The absolute difference of tensor elements between the neighbour steps in the case of LN-TNR (b) and Loop-TNR (d). (b) and (d) are plotted using the same gray scale.}
\label{fig:FixPoint}
\end{figure}

As shown in Fig.~\ref{fig:SquareSym}(d), the whole tensor network is made up by a single $\mathbf{S}$ tensor. Since the $\mathbf{S}$ tensor is rank $3$, we plot the absolute value of the individual tensor elements as a 3D scatter graph in Fig.~\ref{fig:FixPoint}(a) and (c). The $x,y,z$ axes denote the indices of the tensor, and the opacity and color of the scatters denote the amplitude of the tensor elements. The darker means the larger amplitude. We plot several steps when the central charge and scaling dimensions remain accurate. In the case of LN-TNR, we plot from the $5$th to the $9$th step. In the case of Loop-TNR, we plot from the $14$-th to the $18$-th step. In Fig.~\ref{fig:FixPoint}(b) and (d), we show the absolute difference of tensor elements between the $i$th and the $(i+1)$th step under the same gray level. When using LN-TNR, the low-index parts of the tensor are approximately invariant, but the high-index parts is significantly changed. So that only the low-index parts can be regarded as a fixed point tensor. When using Loop-TNR, all tensor elements are nearly invariant. We approximately recover the scale invariance at criticality using Loop-TNR.
\end{document}